\documentclass[useAMS,usenatbib]{mn2e}
\usepackage{graphicx}

\usepackage{times}

\def \aap{A\&A}
\def \aaps{A\&AS}
\def \aj{AJ}
\def \apj{ApJ}

\def \asp{Astron. Soc. Pac.}
\def \baas{BAAS}
\def \jqsrt{J. Quant. Spectrosc. Radiat. Transfer}
\def \mnras{MNRAS}
\def \nat{Nat}
\def \prl{Phys. Rev. Lett.}

\def \da{\Delta\alpha/\alpha}
\def\bsp{\vspace{0.5cm}\small\noindent This paper
has been typeset from a \TeX / \LaTeX\ file prepared by the author.}

\title[Search for high-$z$ molecular absorption lines toward
  millimetre-loud, optically faint quasars]{A search for high redshift
  molecular absorption lines toward millimetre-loud, optically faint
  quasars}
\author[M. T. Murphy, S. J. Curran \&
  J. K. Webb]{M. T. Murphy,$^{1,2}$\thanks{E-mail: mim@ast.cam.ac.uk (MTM)}
  S. J. Curran$^2$ and J. K. Webb$^2$\\
$^{1}$Institute of Astronomy, University of Cambridge, Madingley Road,
  Cambridge, CB3 0HA, UK\\
$^{2}$School of Physics, University of New South Wales, Sydney N.S.W. 2052,
  Australia}

\begin{document}

\date{Accepted ---. Received ---; in original form ---}

\pagerange{\pageref{firstpage}--\pageref{lastpage}} \pubyear{2003}

\maketitle

\label{firstpage}

\begin{abstract}
We describe initial results of a search for redshifted molecular absorption
toward four millimetre-loud, optically faint quasars. A wide frequency
bandwidth of up to 23\,GHz per quasar was scanned using the Swedish-ESO
Sub-millimetre Telescope at La Silla. Using a search list of commonly
detected molecules, we obtained nearly complete redshift coverage up to
$z_{\rm abs}=5$. The sensitivity of our data is adequate to have revealed
absorption systems with characteristics similar to those seen in the four
known redshifted millimetre-band absorption systems, but none were
found. Our frequency-scan technique nevertheless demonstrates the value of
wide-band correlator instruments for searches such as these. We suggest
that a somewhat larger sample of similar observations should lead to the
discovery of new millimetre-band absorption systems.
\end{abstract}


\section{Introduction}\label{sec:intro}
Millimetre-band (mm-band) molecular absorption systems along the
line-of-sight to quasars provide a powerful probe of cold gas in the early
Universe. \citet{WiklindT_94b, WiklindT_95a, WiklindT_96b} have used
molecular absorption lines to study a variety of properties of the
absorbers themselves (e.g.~relative column densities, kinetic and
excitation temperatures, filling factors etc.). Besides information about
the absorbers, important cosmological parameters can be extracted from such
data. Constraints on the cosmic microwave background temperature can be
obtained by comparing the optical depths of different rotational
transitions [e.g.~CO(1--2), CO(2--3) etc.]
\citep[e.g.][]{WiklindT_96a}. Also, if the background quasar is
gravitationally lensed, time delay studies can yield constraints on the
Hubble constant \citep[e.g.][]{WiklindT_01a}. However, these studies have
so far been limited by the paucity of mm-band molecular absorbers. Only 4
such systems are currently known: $z_{\rm abs}=0.685$ toward TXS
0218$+$357 \citep{WiklindT_95a}, $z_{\rm abs}=0.247$ toward PKS 1413$+$135
\citep{WiklindT_97b}, $z_{\rm abs}=0.672$ toward TXS 1504$+$377
\citep{WiklindT_96c} and $z_{\rm abs}=0.886$ toward PKS 1830$-$211
\citep{WiklindT_98a}.

Quasar absorption lines can also be used to search for possible variations
in the fundamental constants. Detailed studies of the relative positions of
heavy element optical transitions in 49 high redshift ($0.5 < z_{\rm abs} <
3.5$) absorption systems favour a smaller fine structure constant ($\alpha
\equiv e^2/\hbar c$) at the 4.1\,$\sigma$ significance level
\citep{MurphyM_01a,WebbJ_01a}. The observed fractional change in $\alpha$
[$\da = (-0.72 \pm 0.18)\times 10^{-5}$] is very small and systematic
errors have to be carefully considered. However, a thorough search for
systematics has not revealed a simpler explanation of the optical results
\citep{MurphyM_01b}. Independent constraints at similar redshifts are
required and recent attention has focused on molecular absorption systems.

Comparison of molecular rotational (i.e.~mm-band) and corresponding H{\sc
\,i} 21-cm absorption line frequencies has the potential to constrain
changes in $\alpha$ with a fractional precision $\sim\!10^{-6}$ per
absorption system -- an order of magnitude gain per absorption system over
the purely optical methods. The ratio of the hyperfine (21-cm) transition
frequency to that of a molecular rotational line is $\propto y \equiv
\alpha^2g_p$ for $g_p$ the proton $g$-factor \citep{DrinkwaterM_98a}. Thus,
any variation in $y$ will be observed as a difference in the apparent
redshifts, $\Delta y/y \approx
\Delta z/(1+z) = (z_{\rm mol} - z_{\rm H})/(1+z_{\rm
mol}$). \citet{CarilliC_00a} and
\citet{MurphyM_01d} have obtained constraints on $\Delta y/y$ consistent
with zero $y$-variation from spectra of PKS 1413$+$135 and TXS
0218$+$357. Currently, the major uncertainty in this mm/H{\sc \,i} comparison
is that intrinsic velocity differences between the mm and H{\sc \,i}
absorption lines are introduced if the lines-of-sight to the mm and radio
continuum emission regions of the quasar differ, as is certainly the case for
PKS 1413$+$135 and TXS 0218$+$357 \citep{CarilliC_00a}.

\begin{table*}
\centering
\begin{minipage}{16.7cm}
\caption{The source list.  $S_{1.4}$, $S_{2.7}$,...  $S_{230}$ are the
measured 1.4, 2.7,... 230 GHz continuum flux densities in Jy. All values
are obtained from the current version of the Parkes catalogue
\citep{WrightA_90a}, except $^a$\citet{PhillipsR_95a}, the $^b$NVSS
catalogue \citep{CondonJ_98a}, $^c$\citet{TeraesrantaH_98a},
$^d$\citet{KovalevY_99a} and $^e$\citet*{TornikoskiM_00a}. No optical
magnitudes are published for these objects. However, note that the
sight-lines to both B 0648$-$165 and B 0727$-$115 suffer from moderate
Galactic extinction, $A_B$ \citep*{SchlegelD_98a}. Also, B 0500$+$019 is
known to have an intervening H{\sc \,i} absorption system. Since we do not
reliably detect continuum emission from this quasar (see
Fig.~\ref{fig:fluxes}), we cannot place interesting limits on HCO$^+$(1--2)
or CO(2--3) absorption at 112.570\,GHz and 218.227\,GHz respectively.}
\label{source}
\begin{tabular}{lcrclllccccccc}\hline
Quasar  &\multicolumn{2}{c}{Coordinates (J2000)}       &$A_B$ (mag)&\multicolumn{8}{c}{Radio flux densities (Jy)} \\
                           & h ~m ~s  & d~~~$'$~~~$''$&           &$S_{1.4}$&$S_{2.7}$&$S_{5.0}$&$S_{8.4}$&$S_{11}$&$S_{22}$&$S_{86}$&$S_{230}$           \\\hline
B 0500$+$019/J 0503$+$203  &05 03 21.2& 02 03 05      &0.289      &2.10     &2.46    &2.04      &1.61     &1.36$^d$&0.86$^d$&--      & --                 \\
B 0648$-$165/J 0650$-$1637 &06 50 24.6&-16 37 40      &2.456      &1.70     &1.40    &1.02      &0.80     &--      &--      &0.9$^a$ & --                 \\
B 0727$-$115/J 0730$-$1141 &07 30 19.1&-11 41 13      &1.271      &2.66$^b$ &1.95    &2.22      &3.36     &3.87$^d$&2.91$^d$&0.9$^a$ & --                 \\
B 1213$-$172/J 1215$-$1731 &12 15 46.8&-17 31 45      &0.253      &1.50     &1.33    &1.28      &1.56     &2.97$^d$&2.44$^d$&1.5$^a$ & $\approx 0.7^{c,e}$\\\hline
\end{tabular}
\end{minipage}
\end{table*}

A {\it statistical} sample of mm/H{\sc i} comparisons is therefore required
to provide a reliable, independent check on the optical results for
$\alpha$-variation. One systematic approach to finding more mm-band
molecular absorbers is to scan the frequency space toward a sample of
millimetre-loud quasars. Indeed, the absorber toward PKS 1830$-$211 was
identified in this way by \citet{WiklindT_96a}. With the assumption that
molecular absorption will be associated with significant optical
extinction, one should select optically faint quasars to increase the
probability of detecting molecular absorption. In this paper we present
wide-band millimetre-wave spectra of the four millimetre-loud quasars which
have not yet been optically identified: non-detections in the
APM\footnote{Available at http://www.ast.cam.ac.uk/$\sim$apmcat} and
DSS\footnote{Available at http://archive.stsci.edu/dss} catalogues imply
$m_B > 20$. In the following section we describe the observations and data
reduction and present the wide-band spectra. In Section \ref{sec:abs} we
search for possible millimetre absorption systems close to our detection
limits. We make our conclusions and discuss the future of our search
technique in Section
\ref{sec:disc}.

\section{Quasar spectra}\label{sec:data}

\subsection{Observations}\label{subsec:obs}

We observed the quasars listed in Table \ref{source} in February 2002 with
the 15-m SEST at La Silla, Chile. The receivers were tuned to
single-sideband mode and typical system temperatures, on the $T_A^*$-scale,
were $\approx$250\,K for the SESIS RX100 and RX150 receivers and
$\approx$340\,K and 480\,K for the IRAM RX115 and RX230 respectively. The
backends were acousto-optic spectrometers with 1440 channels and a channel
width of 0.7\,MHz. We used dual-beam switching with a throw of
$\approx$12\arcmin\ in azimuth, and pointing errors were typically
$3\arcsec$ rms on each axis.

We used the SESIS and IRAM RX230 receivers for the majority of our
integrations, providing the advantage of a wide (1\,GHz) bandwidth (the
IRAM RX115 has a 0.5\,GHz bandwidth). Since the backend response decreases
sharply toward the edges of the band, we overlapped the bands by observing
at intervals of 0.8\,GHz to ensure uniform signal-to-noise ratio (S/N) in
the final, combined spectra. The lowest frequency we could tune the RX100
receiver to was 78.8\,GHz and, although the nominal range is 78--116\,GHz,
the backend configuration did not allow tuning to 81.2, 82.0 or
82.8\,GHz. Similarly, we could not tune the RX150 receiver to 141.7 or
142.5\,GHz.

\subsection{Data reduction}\label{subsec:red}

\begin{figure}
\centerline{\includegraphics[width=8.4cm]{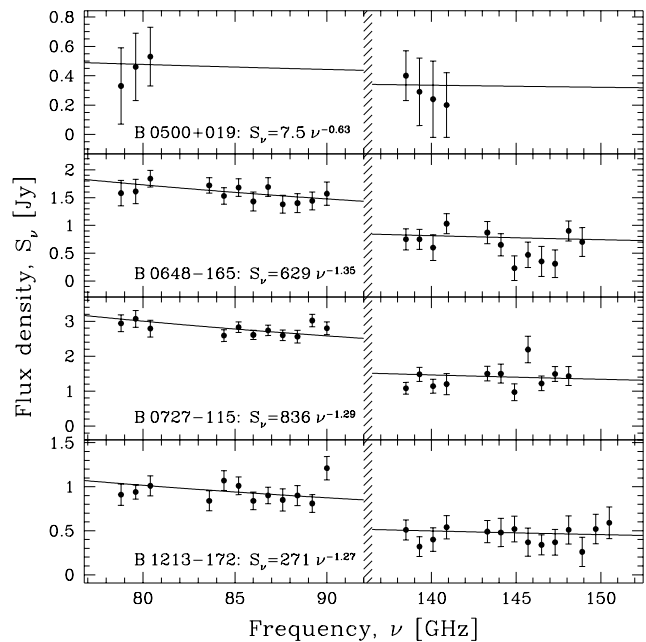}}
\caption{The average flux densities, $S_\nu$, for each 1\,GHz-wide portion of
spectrum. The error bars represent the rms derived from typically 10 scans
at each frequency, $\nu$. Power-law fits with the given parameters were
used to define the continuum flux to which the spectra in
Figs.~\ref{fig:0500}--\ref{fig:1213} have been normalized.}
\label{fig:fluxes}
\end{figure}

For each 1\,GHz-wide spectrum, we averaged typically 10 two-minute
(reference-subtracted) scans of the source and subtracted a low order
polynomial fit from the data to provide a flat continuum. To convert the
data from the $T_A^*$-scale to the Jy-scale we fitted a third-order
polynomial to the aperture efficiency of SEST as a function of frequency:
\begin{eqnarray}\label{eq:eff}
S_\nu/{\rm Jy} & = & 7.55 + 0.340(\nu/{\rm GHz}) - 2.07\times
10^{-3}(\nu/{\rm GHz})^2 \nonumber \\ & & + 5.32\times 10^{-6}(\nu/{\rm
GHz})^3\,.
\end{eqnarray}
The intensity was calibrated using the chopper-wheel method which should be
accurate to $\pm$10\%\footnote{From the SEST handbook,
http://www.ls.eso.org/lasilla/Telescopes/SEST}. However, we found
significant variations in the total flux density between each two-minute
scan. Therefore, in Fig.~\ref{fig:fluxes} we present the mean flux density,
$S_\nu$, for each of the 1\,GHz-wide spectra with error bars representing
the rms from the contributing scans.

\begin{figure*}
\vspace{0.7cm}
\centerline{\includegraphics[height=17cm,width=10cm,angle=270]{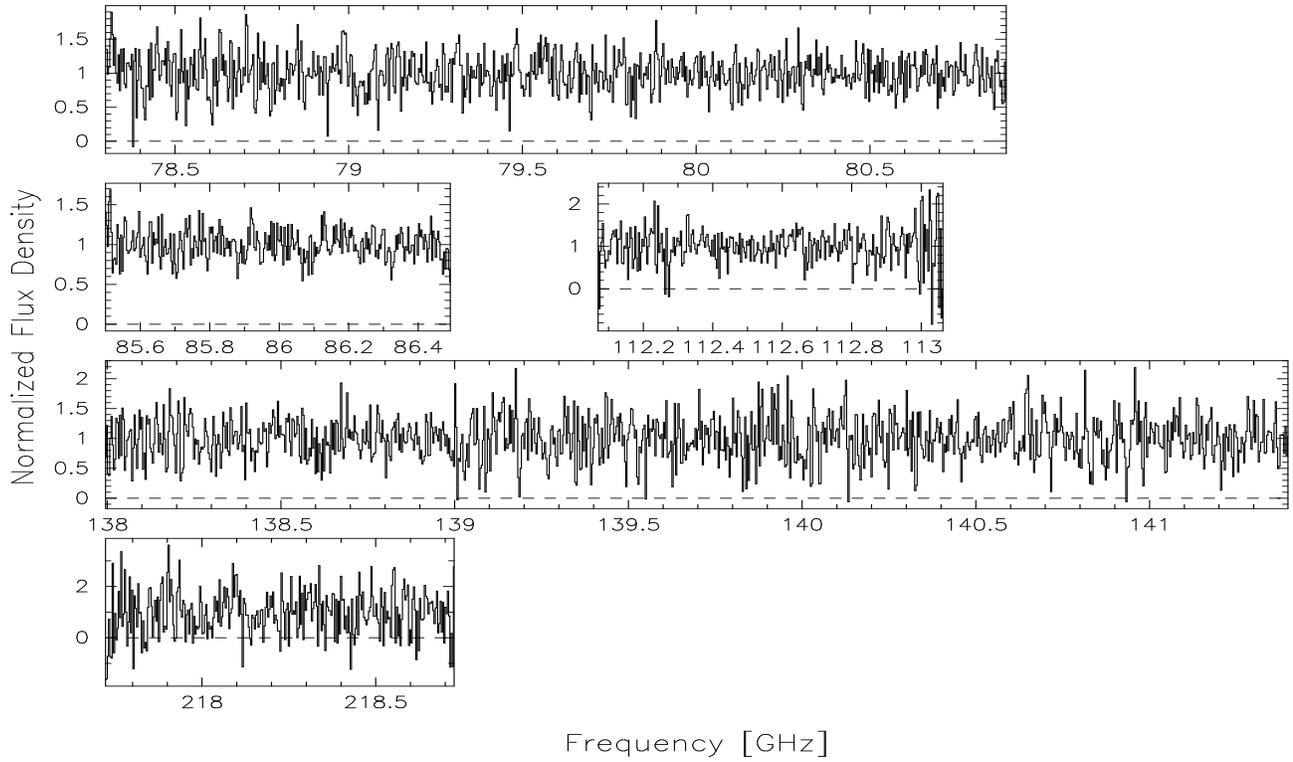}}
\caption{Wide-band spectrum of B 0500$+$019. These data (and those in
Figs.~\ref{fig:0648}--\ref{fig:1213}) are normalized by the continuum flux
of the quasar (taken as a power-law fit to the flux density data in
Fig.~\ref{fig:fluxes}) and, for clarity, have been boxcar-smoothed over 3
channels. The mean S/N = 1.72 per (unsmoothed) channel.}
\label{fig:0500}
\end{figure*}

\begin{figure*}
\vspace{1.0cm}
\centerline{\includegraphics[height=17cm,width=10cm,angle=270]{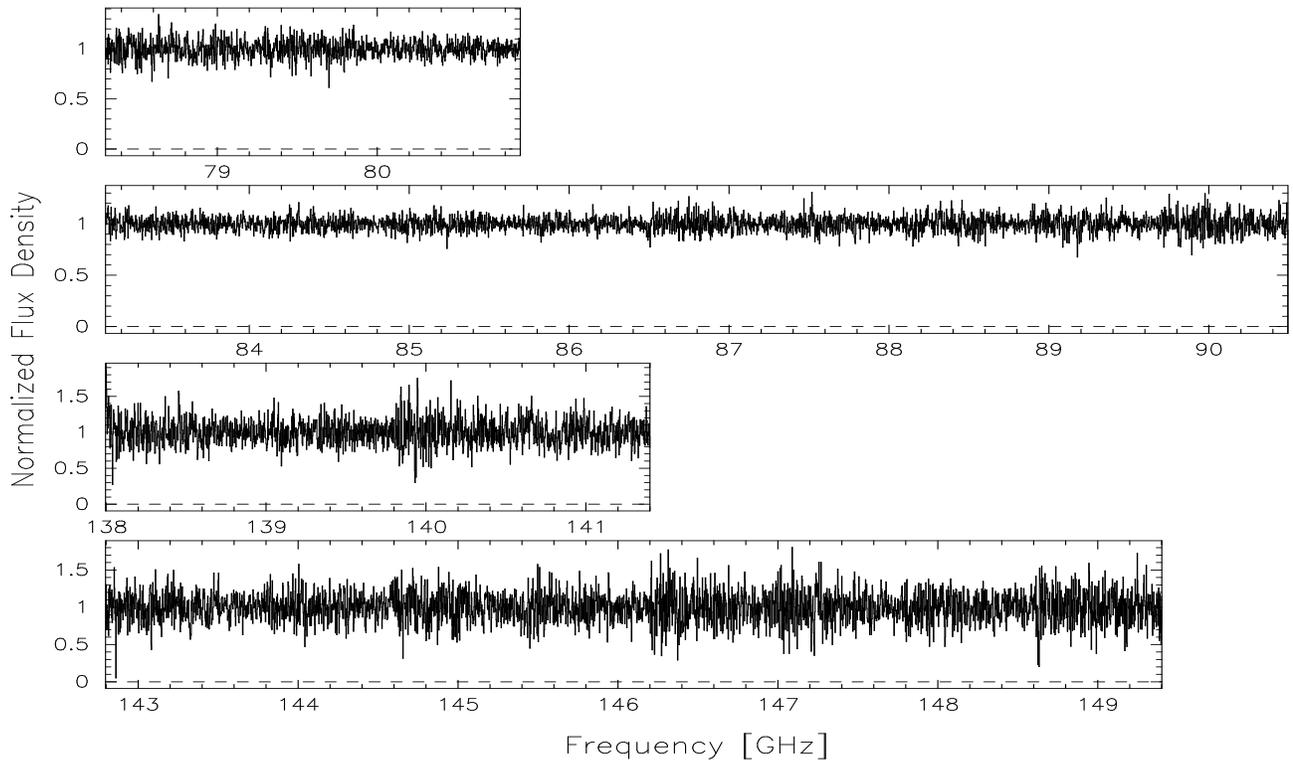}}
\caption{Wide-band spectrum of B 0648$-$165 (see caption to
Fig.~\ref{fig:0500} for details). The mean S/N = 5.73 per (unsmoothed)
channel.}
\label{fig:0648}
\end{figure*}

\begin{figure*}
\vspace{0.7cm}
\centerline{\includegraphics[height=17cm,width=10cm,angle=270]{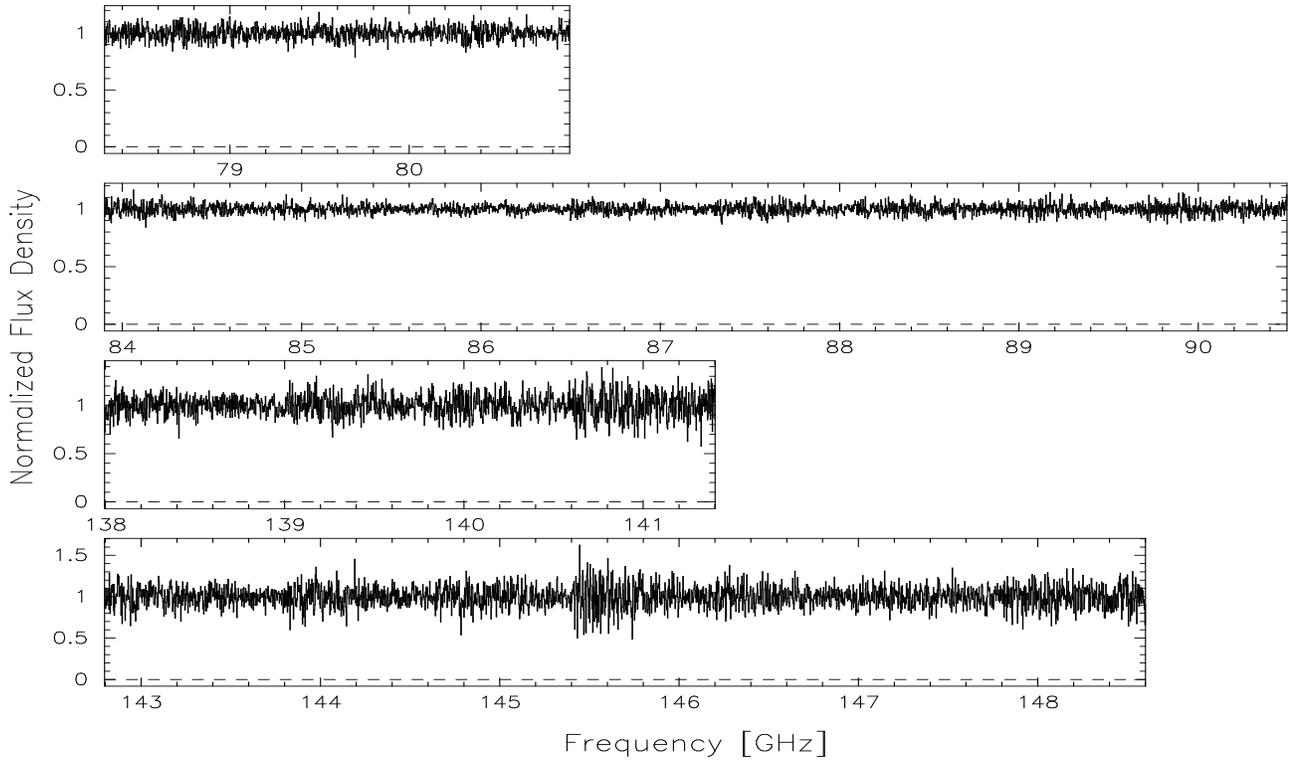}}
\caption{Wide-band spectrum of B 0727$-$115 (see caption to
Fig.~\ref{fig:0500} for details). The mean S/N = 9.63 per (unsmoothed)
channel.}
\label{fig:0727}
\end{figure*}

\begin{figure*}
\vspace{1.0cm}
\centerline{\includegraphics[height=17cm,width=10cm,angle=270]{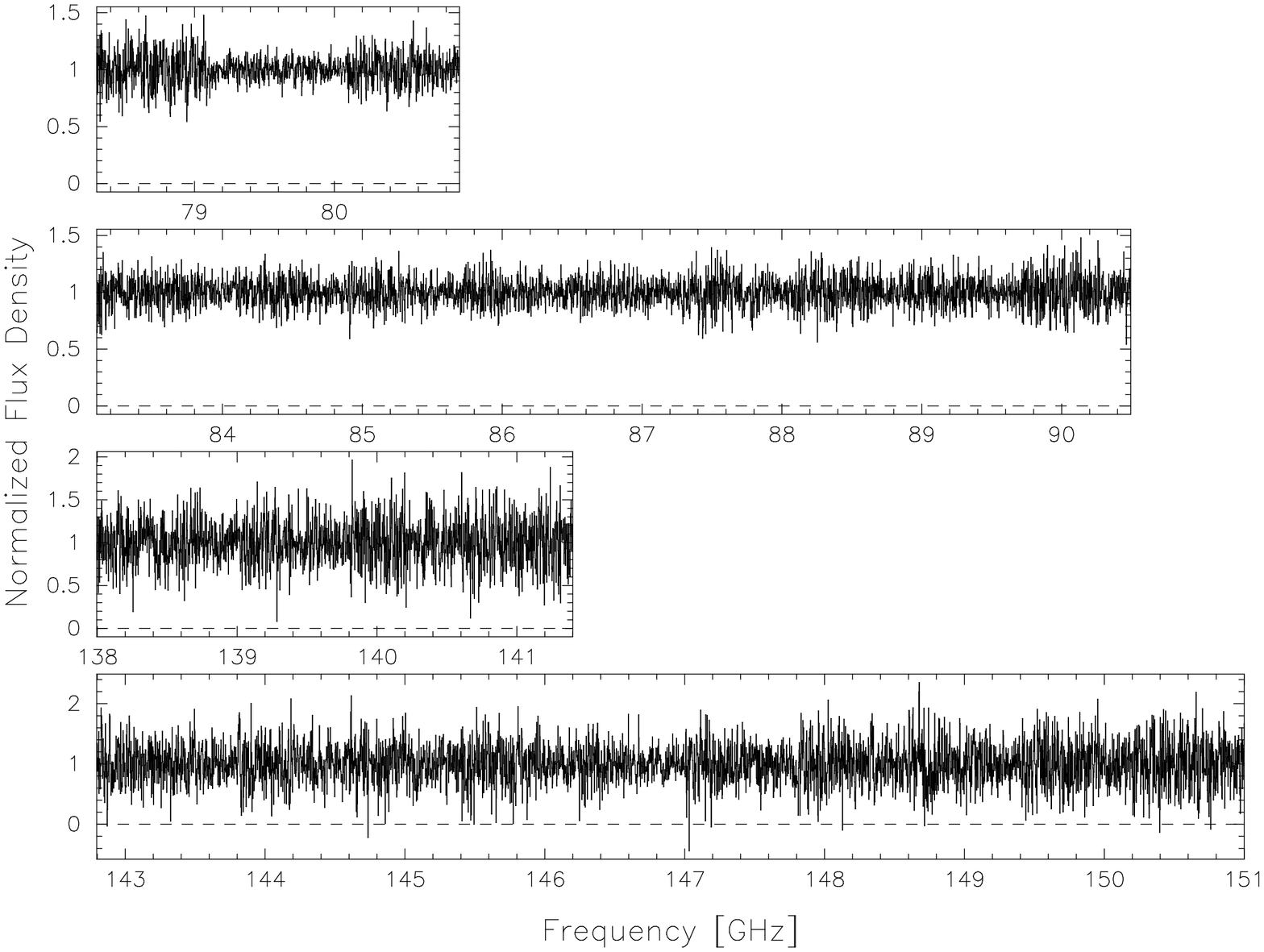}}
\caption{Wide-band spectrum of B 1213$-$172 (see caption to
Fig.~\ref{fig:0500} for details). The mean S/N = 3.40 per (unsmoothed)
channel.}
\label{fig:1213}
\end{figure*}

For each channel, we generated a 1\,$\sigma$ error from the rms in a window
of width $2N_{\rm err}+1$ channels centred on that channel. We found that
$N_{\rm err} = 20$ provided reliable rms values but the results in Section
\ref{subsec:results} were insensitive to this parameter. In the 0.2\,GHz
regions where different 1\,GHz-wide spectra overlapped, we re-sampled the
variance-weighted average using the largest channel width of the
contributing spectra. Finally, the combined spectra and 1\,$\sigma$ error
arrays were normalized to the flux density, $S_\nu$, of the
quasars. $S_\nu$ was taken as a power-law fit to the flux densities
presented in Fig.~\ref{fig:fluxes}.

The combined spectra, normalized by the continuum, are presented in
Figs.~\ref{fig:0500}--\ref{fig:1213}. Each contiguous spectral segment is
plotted separately and all regions are plotted with the same linear
frequency-scale. For clarity, the data have been boxcar-smoothed over 3
channels (we analyse the unsmoothed spectra in Section \ref{sec:abs}). From
Fig.~\ref{fig:0500} it is clear that we only barely detect the continuum of
B 0500$+$019. However, B 0648$-$165, B 0727$-$115 and B 1213$-$172 have S/N
as high as $\approx$10 per (unsmoothed) 0.7\,MHz channel. We obtained the
largest spectral coverage on B 1213$-$172, a total (discontinuous)
bandwidth of 23.2\,GHz.

\section{Searching for absorption systems}\label{sec:abs}

\subsection{Search algorithm}\label{subsec:algor}

Table \ref{red} lists the transitions detected in the 4 known mm-band
molecular absorption systems. From this table we selected a set of
`commonly' detected molecules for which to search in our observed spectra:
CO, HCO$^+$, HCN, HNC, CS and CN. We also searched for some common
isotopomers of these molecules, which have lower terrestrial abundances, to
provide potentially greater redshift space coverage and, as explained
below, to help rule out some candidate absorption systems: $^{13}$CO,
C$^{18}$O, C$^{17}$O, H$^{13}$CO$^+$, H$^{13}$CN. For each molecule, we
searched for all transitions lying below 1000\,GHz (rest-frame) listed in
the molecular line database of \citet{PickettH_98a}\footnote{Available at
http://spec.jpl.nasa.gov}. With this fiducial set of transitions we obtain
complete redshift coverage up to $z\approx 10$ for B 0648$-$165, B
0727$-$115 and B 1213$-$172. Since the frequency coverage for B 0500$+$019
is significantly smaller, several small redshift gaps (i.e.~$\Delta z \la
0.05$) appear for $0.19 < z < 3.84$. However, the fraction of redshift
space covered over this range is still high, $\approx$85\%. The redshift
coverage is complete for all other redshifts below $z\approx 10$.

\begin{table}
\centering
\begin{minipage}{7.0cm}
\caption{Transitions already detected in absorption at high redshift.}
\label{red}
\begin{tabular}{l l l}\hline
Transition                     & Quasar$^a$& Reference$^b$      \\\hline
CO(0--1)                       & B         & WC94b              \\
CO(1--2)                       & A,C       & WC95; WC96a        \\
CO(2--3)                       & A,C       & WC95; WC96a        \\
CO(3--4)                       & D         & WC98               \\
$^{13}$CO(1--2)                & A,C       & WC95; WC96a        \\
C$^{18}$O(1--2)                & A         & CW95               \\ \\
HCO$^+$(0--1)                  & B,D       & WC97; CM02         \\
HCO$^+$(1--2)                  & A,B,C,D   & WC95; WC96a,b,c    \\
HCO$^+$(2--3)                  & B,C,D     & WC96a; WC96c; WC97 \\
HCO$^+$(3--4)                  & C         & WC96a              \\
H$^{13}$CO$^+$(0--1)           & D         & CM02               \\
H$^{13}$CO$^+$(1--2)           & D         & WC98               \\ \\
HCN(1--2)                      & A,B,C,D   & WC95; WC96a,b; WC97\\
HCN(2--3)                      & B,D       & WC96c; WC97        \\ \\
HC$_3$N(4--5)                  & D         & CM02               \\ \\
H$^{13}$CN(1--2)               & D         & WC98               \\ \\
HNC(0--1)                      & D         & CM02               \\
HNC(1--2)                      & B,C,D     & WC96a,c; WC97      \\
HNC(2--3)                      & B,C,D     & WC96a,c; WC97      \\ \\
CS(0--1)                       & A         & CWN97              \\
CS(2--3)                       & D         & WC96c              \\
CS(3--4)                       & D         & WC96c              \\ \\
N$_2$H$^+$(1--2)               & D         & WC96c              \\
N$_2$H$^+$(2--3)               & D         & WC96c              \\ \\
H$_{2}$CO($1_{10}$--$2_{11}$)  & D         & WC98               \\
\\
H$_{2}$O(ortho)                & A       & CW98a \\
\\
C$_3$H$_2$($1_{01}$--$2_{12}$) & D       & CM02\\
\\
LiH(0--1)                      & A       & CW98b\\\hline
\end{tabular}
\footnotesize{$^a$Quasars: A = TXS 0218$+$357, B = PKS 1413$+$135, C = TXS
1504$+$377, D = PKS 1830$-$211. $^b$References: CM02 \citep{CarilliC_02a},
CW95 \citep{CombesF_95a}, CWN97
\citep*{CombesF_97a}, CW98a \citep{CombesF_98c}, CW98b \citep{CombesF_98a},
WC94b \citep{WiklindT_94a}, WC95 \citep{WiklindT_95a}, WC96a
\citep{WiklindT_96c}, WC96b
\citep{WiklindT_96b}, WC96c \citep{WiklindT_96a}, WC97
\citep{WiklindT_97b}, WC98 \citep{WiklindT_98a}.}
\end{minipage}
\end{table}

\begin{figure}
\centerline{\includegraphics[width=8.4cm]{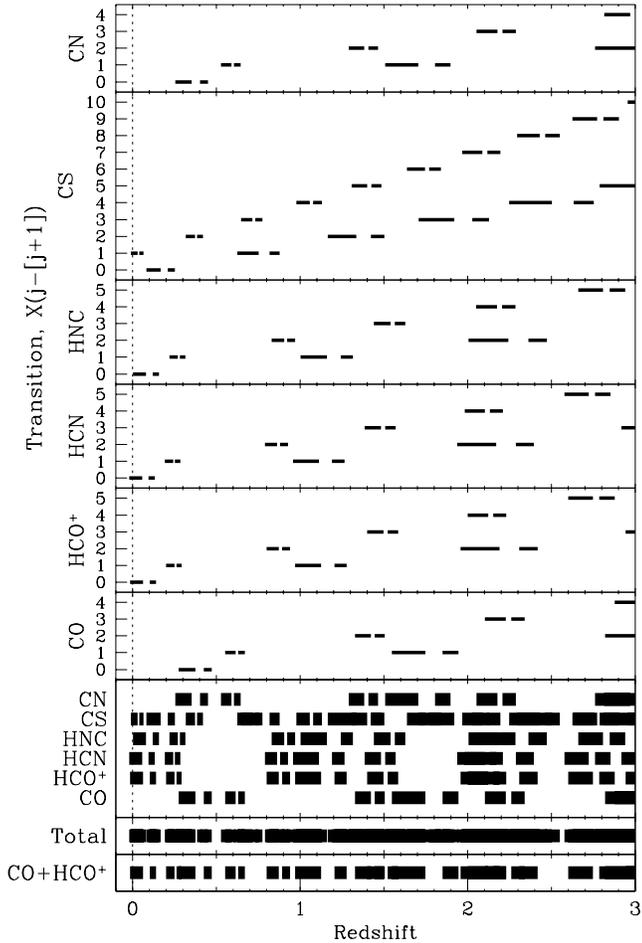}}
\caption{Schematic diagram illustrating the redshift coverage for B
1213$-$172 using only the strong (i.e. highest terrestrial abundance)
isotopomers in the fiducial set. The upper six panels show the redshifts
covered by each molecule given the frequencies observed (see
Fig.~\ref{fig:1213}). The vertical scale shows the lower-level rotational
quantum number, $j$. The third panel from the bottom summarizes the
coverage from all $j$-levels for each molecule. The panel marked `Total'
shows the coverage from all molecules and the lowest panel includes only
the coverage from CO and HCO$^+$. See text for discussion.}
\label{fig:coverage}
\end{figure}

In Fig.~\ref{fig:coverage} we illustrate the redshift coverage up to
$z_{\rm abs}=3.0$ achieved for B 1213$-$172 using only the strong
(i.e. highest terrestrial abundance) isotopomers in the fiducial set. From
the references in Table \ref{red} it is clear that the transitions of HCN,
HNC, CS and CN typically present a lower optical depth for absorption
compared to CO and HCO$^+$. Therefore, our estimate of the redshift range
covered using the fiducial set of transitions may appear optimistic. The
lower two panels in Fig.~\ref{fig:coverage} compare the redshift coverage
when using all the strong isotopomers with that obtained using only CO and
HCO$^+$. Though a noticeable decrease in the redshift coverage is clear,
just using CO and HCO$^+$ as a search diagnostic still allows much of the
redshift space to be scanned. Note, however, that the third panel (from the
bottom) in Fig.~\ref{fig:coverage} shows that the redshifts covered by CO
rarely overlap with those covered by HCO$^+$. Thus, if one is to identify
absorption {\it systems} with such a redshift scanning technique, one is
biased toward finding systems containing strong HCN, HNC and perhaps CS and
CN. This bias can only be avoided by observing wider-band spectra.

Using the 1\,$\sigma$ error array for each spectrum, we slid a window of
width $N_{\rm win}$ along the spectrum to search for absorption features
above a significance level of $N_\sigma$ standard deviations. Note that the
apparent significance of a candidate feature will be influenced by any
correlations between the flux in adjacent channels. We address this problem
in the simulations described in Section \ref{subsec:results}. Once an
absorption feature is identified we associate it with a particular
transition, allowing us to assign to it an absorption redshift, $z_{\rm
abs}$, with an appropriate error, $\delta z_{\rm abs}$, which is taken as
the redshift interval corresponding to the half-width of the feature. The
`width' is defined here as the number of channels over which the feature
remains significant at a level $\ge N_\sigma$ standard deviations.

To identify an absorption {\it system} we search for another absorption
feature which, when associated with a different transition, has a redshift
$z_{\rm abs}\pm \delta z_{\rm abs}$. For computational convenience we
impose an arbitrary but conservative redshift upper limit for absorption
systems, $z_{\rm abs} \le 5$. This upper limit can be extended to higher
redshifts but our conclusions below are unchanged. Having identified a
potential absorption system we can find the significance of absorption for
all other transitions in the set defined above which fall within the
bandpass of the spectrum.

Some of the candidate absorption systems can be immediately rejected using
self-consistency conditions imposed by the relative `strengths' of all
available transitions. For example, if the detection of the absorption
system relies on significant absorption in $^{13}$CO(1--2) but there is no
significant absorption in CO(1--2) then we can rule out the putative
detection. To form a rigid selection criteria, we found the integrated flux
density, $S_{\rm int}$, and 1\,$\sigma$ error, $\delta S_{\rm int}$, for
the `host' line [i.e.~the weaker but higher terrestrial abundance line;
CO(1--2) in the example above] and the significantly `detected' line. In
the analysis below we ruled out candidate systems if
\begin{equation}\label{eq:select}
S_{\rm int}^{\rm host} > S_{\rm int}^{\rm det} + 2\,(\delta S_{\rm
  int}^{\rm host} + \delta S_{\rm int}^{\rm det})\,.
\end{equation}
With the target transitions selected above, the fraction of absorption
systems rejected in this way was typically $\sim$20\%. Relaxing the
criteria in equation \ref{eq:select} does not alter our main conclusions
below.

\subsection{Search results}\label{subsec:results}

Consistent with a visual inspection of Figs.~2--5, no strong
(i.e. $\tau\sim 1$) absorption systems exist in our data. For all 4 quasars
in our sample, no single-line absorption features were identified with
$N_\sigma \ge 6.5\,\sigma$ for $3 \le N_{\rm win} \le 31$. No absorption
{\it systems} (i.e. double-line features) were observed for $N_\sigma \ge
5.4\,\sigma$. We list the strongest absorption {\it system} candidates in
Table \ref{tab:cand}.

\begin{table}
\centering
\begin{minipage}{7.3cm}
\caption{The most significant double-line detections for each quasar in our
sample. All transitions which fell into the band-pass of our observations
are given and column 4 shows the significance of each detection (not
corrected for correlations between data channels; negative values indicate
flux values above the continuum). The final column gives the width in
channels (as defined in Section \ref{subsec:algor}), $\Delta$, of the two
significant features responsible for the putative detection.}
\label{tab:cand}
\begin{tabular}{cclrc}\hline
Quasar      &$z_{\rm abs}$&Transitions         &Sig. ($\sigma$)&$\Delta$\\\hline
B 0500$+$019&2.81483      &CS(10--11)          &$ 4.5$         &2       \\
            &             &CS(16--17)          &$-1.0$         &        \\
            &             &C$^{18}$O(2--3)     &$ 4.8$         &2       \\
            &             &HCO$^+$(5--6)       &$-2.1$         &        \\
            &             &HCN(5--6)           &$ 0.6$         &        \\ \\
B 0648$-$165&2.02869      &HCO$^+$(4--5)       &$ 4.3$         &2       \\
            &             &HCO$^+$(2--3)       &$ 0.8$         &        \\
            &             &H$^{13}$CO$^+$(2--3)&$ 0.4$         &        \\
            &             &H$^{13}$CO$^+$(4--5)&$-0.8$         &        \\
            &             &$^{13}$CO(3--4)     &$ 0.9$         &        \\
            &             &C$^{17}$O(3--4)     &$ 0.0$         &        \\
            &             &C$^{18}$O(3--4)     &$ 1.9$         &        \\
            &             &HCN(2--3)           &$ 5.4$         &5       \\
            &             &HCN(4--5)           &$ 0.0$         &        \\
            &             &HNC(2--3)           &$ 0.3$         &        \\
            &             &CS(4--5)            &$ 0.1$         &        \\
            &             &CS(8--9)            &$ 0.9$         &        \\ \\
B 0727$-$115&1.51374      &CN(1--2)            &$ 4.2$         &2       \\
            &             &$^{13}$CO(1--2)     &$ 0.1$         &        \\
            &             &C$^{17}$O(1--2)     &$ 3.0$         &        \\
            &             &C$^{18}$O(1--2)     &$ 4.9$         &4       \\
            &             &H$^{13}$CO$^+$(3--4)&$ 0.4$         &        \\
            &             &HCN(3--4)           &$-1.3$         &        \\
            &             &HNC(3--4)           &$ 0.8$         &        \\ \\
B 1213$-$172&1.81705      &C$^{17}$O(1--2)     &$ 4.4$         &2       \\
            &             &CN(1--2)            &$ 5.0$         &4       \\
            &             &CS(4--5)            &$ 2.1$         &        \\
            &             &CS(7--8)            &$ 1.4$         &        \\\hline
\end{tabular}
\end{minipage}
\end{table}

\begin{figure}
\centerline{\includegraphics[width=8.4cm]{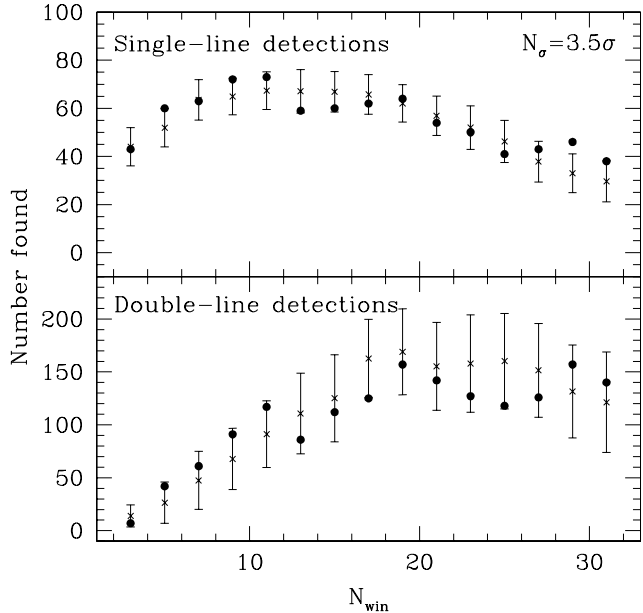}}
\caption{Number of single- and double-line matches found for B 1213$-$172
(solid circles) compared with the number found in simulated data (crosses)
for a range of $N_{\rm win}$ and $N_\sigma=3.5\,\sigma$. The 1\,$\sigma$
error bars are derived from the rms of 200 simulations and the crosses
represent the mean value.}
\label{fig:compsim}
\end{figure}

Our search technique can be used to detect weaker absorption lines/systems
by lowering the rejection limit, $N_\sigma$. However, this results in
detection of large numbers of single and double-line features. For example,
Fig.~\ref{fig:compsim} shows the number of features detected in B
1213$-$172 if we set $N_\sigma = 3.5\,\sigma$ (solid circles). To be sure
that most of the these `weak candidates' are spurious (i.e. the result of
noise) we constructed synthetic spectra with the following procedure:
\begin{enumerate}
\item For each quasar, we modelled each 1\,GHz integration as Gaussian
  noise with rms per channel equal to that of the real quasar data
  (i.e. the error arrays for the real and synthetic spectra were
  identical).

\item Each synthetic spectrum was convolved with a Gaussian with FWHM equal
  to that of the autocorrelation function of the quasar data.

\item We produced a final combined spectrum using the same procedure used
  for the real quasar spectra (see Section \ref{subsec:red}).
\end{enumerate}
The effect of step (ii) is to introduce positive correlations between the
flux density in neighbouring channels. We confirmed that the amplitude and
range of correlations match those found in the real quasar data by
comparing the autocorrelation functions for both data sets.

In Fig.~\ref{fig:compsim} we compare the number of single- and double-line
detections in our spectrum of B 1213$-$172 with a Monte Carlo simulation
using synthetic spectra produced with the above procedure. We note that the
number of lines identified in the real and synthetic spectra are the same
within the standard deviation of the simulations. We verified that this is
true for the entire range $3 < N_{\rm win} < 31$ (i.e. line widths $5{\rm
\,km\,s}^{-1} < \Delta v < 80{\rm \,km\,s}^{-1}$) and for all rejection limits
$3.0\,\sigma < N_\sigma < 5.5\,\sigma$, suggesting that our simple model of
the SEST data is adequate. Thus, we conclude that we have not detected a
large number of weak absorption systems in the quasar data, though we
cannot rule out single [or small numbers ($\sim$5) of] weak systems.

\section{Discussion}\label{sec:disc}

We have selected 4 millimetre-loud quasars which have not been detected
optically ($m_B > 20$), possibly because an intervening absorption system
causes a large visual extinction. Such dusty systems could lead to
detectable mm-band absorption and so we have performed wide-band
millimetre-wave spectral scans to search for absorption at arbitrary,
unknown redshifts. After defining a set of commonly detected molecules, we
searched for absorption with nearly complete redshift coverage up to
$z_{\rm abs} = 5$. No candidate absorption systems (i.e. two independent
features attributable to transitions with a common redshift) could be
identified where both features had a significance $>5.4\,\sigma$.

Simulations indicate that systems identified with lower significance are
consistent with noise. However, as a reminder that we cannot completely
rule out these putative detections, we provide the best absorption system
candidates in Table \ref{tab:cand}. These can be regarded as priority
targets for follow-up observations. For our highest S/N spectrum, that of B
0727$-$115, the 3\,$\sigma$ optical depth limit is $\approx$0.3\, per
(unsmoothed) channel. This is comparable to the optical depth limits
reached in searches for various molecules in damped Lyman-$\alpha$
absorption systems (see \citealt{CurranS_02b} and references therein) and
compares well with the optical depths of many transitions in the 4 known
high-$z$ mm-band molecular absorbers (see references in Table
\ref{red}). Therefore, despite our null result, it is clear that wide-band
correlator systems can be used to efficiently search for molecular
absorption systems at unknown redshifts.

\section*{Acknowledgments}
We thank M. Anciaux and M. Lerner for reconfiguring the SEST software,
allowing us to tune to a wide range of awkward frequencies. Thanks also to
the anonymous referee who's comments significantly improved the paper. We
are grateful for financial support from the John Templeton Foundation. SJC
received a UNSW NS Global Fellowship. MTM received a Grant-in-Aid of
Research from the National Academy of Sciences, administered by SigmaXi,
the Scientific Research Society. MTM is also grateful to PPARC for support
at the IoA under the observational rolling grant (PPA/G/O/2000/00039). This
research has made use of the NASA/IPAC Extragalactic Database (NED) which
is operated by the Jet Propulsion Laboratory, California Institute of
Technology, under contract with the National Aeronautics and Space
Administration.


\bsp

\label{lastpage}

\end{document}